\documentclass[preprints,article,submit,pdftex,moreauthors]{Definitions/mdpi} 

\usepackage{graphicx}
\usepackage{overpic}
\usepackage{subcaption}
\preto{\abstractkeywords}{\nolinenumbers}

\firstpage{1} 
\pubvolume{1}
\issuenum{1}
\articlenumber{0}
\pubyear{2026}
\copyrightyear{2026}
\datereceived{ } 
\daterevised{ } 
\dateaccepted{ } 
\datepublished{ } 

\Title{Toward in-situ/operando X-ray absorption spectroscopy and electrochemical characterization of solid oxide fuel cells}

\Author{Renato A. N. de Oliveira \textsuperscript{1}*, 
            Rafael Galiza Yoshimura\textsuperscript{1},
            Liliana Mogni\textsuperscript{2}, 
            Maurico Arce\textsuperscript{2},
            Diego G. Lamas\textsuperscript{3},
            Lucia Toscani\textsuperscript{3},
            Maria Belén Arcentales Vera\textsuperscript{3},
            Esteban Elvis Asto Ramos\textsuperscript{3},
            Magna Monteiro Schaerer\textsuperscript{4},
            Celeste Z.A. Aquino\textsuperscript{4},
            Marcia Carvalho de Abreu Fantini\textsuperscript{5},
            Taofeeq Oladayo Bello\textsuperscript{5},
            Tereza da Silva Martins\textsuperscript{6},
            Danilo Waismann Losito\textsuperscript{6},
            James Moraes de Almeida\textsuperscript{7},
            Valeria Spolon Marangoni\textsuperscript{7},
            Leopoldo Suescunand\textsuperscript{8}, 
            and Santiago Figueroa \textsuperscript{9}* }

\AuthorNames{}

\address{%
\textsuperscript{1} \quad LNLS/CNPEM-Brasil \\
\textsuperscript{2} \quad CAB-Argentina \\
\textsuperscript{3} \quad UNSAM-Argentina \\
\textsuperscript{4} \quad UNA-Paraguay \\
\textsuperscript{5} \quad USP-Brasil \\
\textsuperscript{6} \quad UNESP-Brasil \\
\textsuperscript{7} \quad Ilum-Brasil \\
\textsuperscript{8} \quad UdelaR-Uruguai \\
\textsuperscript{9} \quad Unicamp-Brasil \\
}

\corres{Correspondence: renato.oliveira@lnls.br;santiago.figueroa74@gmail.com}

\abstract{ The focus of the present work is the development of specialized experimental instrumentation compatible with synchrotron characterization for in-situ and operando symmetric intermediate temperature solid oxide fuel cells (IT-SOFC) studies at maximum temperatures of 800\textsuperscript{$\circ$}C , exposed to reducing and oxidizing atmospheres, using fluorescence X-ray absorption spectroscopy (XAS) measurements in combination with electrochemical impedance spectroscopy (EIS) in the multipurpose Quati beamline at CNPEM/SIRIUS synchrotron facility \cite{Santiago2023}. Symmetric IT-SOFC  are gaining importance due to their structural simplicity, as they allow for the use of identical materials on both sides of the fuel cell electrolyte; the anode, and the cathode \cite{Fangjie:2025,Muhammad2022}. The symmetric configuration opens new opportunities for fundamental research of electrode materials and improves the versatility of SOFC electrochemical devices \cite{Fangjie:2025,Muhammad2022}.}

\begin{document}
\nolinenumbers



\section{Introduction}
Fuel cells are electrochemical energy conversion devices that directly transform chemical energy into electrical energy without the need for combustion processes, making them essential technologies for the global energy transition toward sustainable power generation systems \cite{Saddam:2020}. In contrast to conventional thermal devices, fuel cells rely on electrochemical reactions occurring at electrodes separated by an ion-conducting electrolyte, enabling high efficiency and low environmental impact \cite{Saddam:2020}. Fuel cells convert chemical energy directly into electrical energy through a redox reaction \cite{Unicamp2025}. Like thermal devices, they operate continuously as long as fuel and oxidant are supplied. Because of this nature, fuel cells are capable of continuously produce electricity with high conversion efficiency.

Several types of fuel cell technologies exist, each defined primarily by the electrolyte material and operating temperature range \cite{skinnerd:2008}. Examples include polymer electrolyte membrane fuel cells, alkaline fuel cells, and solid oxide fuel cells. Despite differences in materials and temperature regimes, all fuel cell technologies share common characteristics such as high energy conversion efficiency, absence of moving mechanical parts, modular scalability, and low or near-zero pollutant emissions during operation.

Applications of fuel cells extend to a wide technological spectrum, from portable devices and backup power systems to transportation and large-scale distributed electricity generation \cite{Unicamp2025}. These technologies contribute to reducing dependence on fossil fuels, lowering greenhouse gas emissions, improving energy security, and supporting the development of a hydrogen-based energy infrastructure.

The current work focuses on  symmetric Intermediate-Temperature Solid Oxide Fuel Cells \cite{JuanCarlos2011,KejunZhu2022,Muhammad2022,Liliana2016} (IT-SOFC), which operate typically between 500 and 700\textsuperscript{$\circ$}C. Operating in this temperature range reduces thermal degradation, and lowers system costs while maintaining sufficient ionic conductivity and catalytic activity. Because of their symmetric configuration and reversible electrochemical behavior, these devices can also function as Solid Oxide Electrolyzer Cells (SOEC), enabling efficient electrolysis of water \cite{Meng2008}. When powered by renewable electricity, SOEC systems can produce green hydrogen, a key energy carrier in decarbonized energy systems.

Solid oxide fuel cell devices are composed of a cathode, an anode, and a dense ceramic electrolyte. At the cathode, oxygen reduction reactions (ORR) convert molecular oxygen into oxide ions. Perovskite oxides such as lanthanum strontium manganite exhibit excellent electrocatalytic performance for this reaction \cite{Sacanell2017,Diego2017,JACOBSON2012478}. The anode promotes fuel oxidation or hydrogen oxidation reaction (HOR), commonly employing nickel yttria stabilized zirconia cermets capable of hydrogen oxidation and internal reforming of methane-containing fuels \cite{Nigel2017,skinnerd:2008}. The electrolyte is typically a dense ceramics, e. g., yttria stabilized zirconia, which provides high oxide-ion conductivity and chemical stability at elevated temperatures \cite{Unicamp2025}.

To study the mechanisms involved on the SOFC operation, a central component of this work is the development of specialized experimental instrumentation compatible with synchrotron X-ray absorption spectroscopy technique. The instrumentation will leverage the capabilities of the Quati beamline for nanoscale investigation of functional materials \cite{Santiago2023}. The methodology involves experimental engineering cells capable of operating under realistic catalytic and electrochemical conditions while allowing X-ray transmission and precise environmental control. These measurement cells will enable direct observation of catalytic reactions, electrocatalytic processes, and ionic transport phenomena in nanostructured materials during operation, establishing a powerful experimental framework for understanding and optimizing next-generation solid oxide electrochemical technologies.

The cells is designed for research and development conducted at the Quati beamline of the Sirius synchrotron located at LNLS/CNPEM in Campinas \cite{LIU2014}. The fourth-generation synchrotron light source provides exceptional brightness, and energy resolution, enabling nanoscale investigation of functional materials under realistic operational environments. The proposed experimental cells will be compatible with electrochemical impedance spectroscopy (EIS) and X-ray absorption spectroscopy measurements, allowing time-resolved analysis of oxidation states, local coordination environments, and dynamic structural transformations in electrode and catalytic materials during operation.

\section{Materials and Methods}
The proposed electrochemical sample cell architecture and components were designed to ensure symmetry between both sides of the fuel cell membrane, while employing low atomic number (Z) materials on critical components to minimize X-ray absorption due to material parts along the X-ray beam, and low resistance gold-plated contacts for the anode and cathode. Those design details maximize the X-ray signal and facilitate a wider range of experimental conditions, which in turn can improve the reproducibility and reliability of the XAS and EIS measurements.

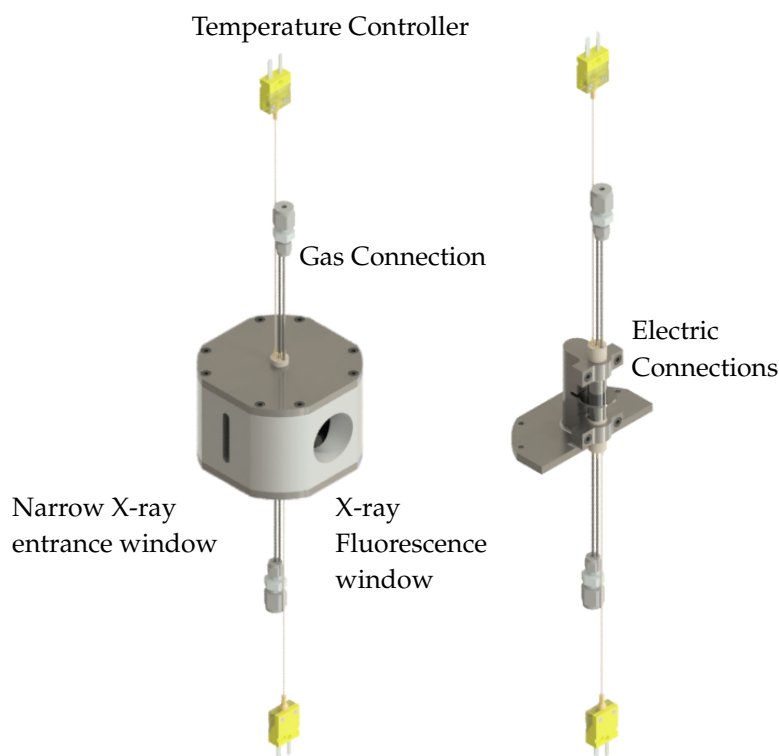
\begin{figure}
     \centering
          \begin{overpic}[width=.15\textwidth]{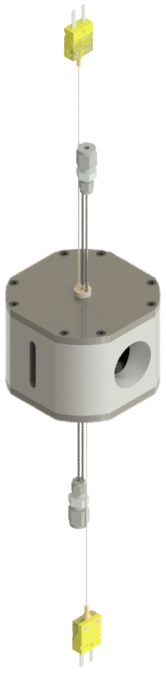}
           \put(0,102){\textcolor{black}{Temperature Controller}}
           
           \put(15,70){\textcolor{black}{Gas Connection}}
           
           \put(20,35){\textcolor{black}{ X-ray }}
           \put(20,30){\textcolor{black}{ Fluorescence}}
           \put(20,25){\textcolor{black}{ window}}

           \put(-25,35){\textcolor{black}{ Narrow X-ray }}
           \put(-25,30){\textcolor{black}{ entrance window}} 
          \end{overpic} \hspace{1.70cm}
         \begin{overpic}[width=0.129\textwidth]{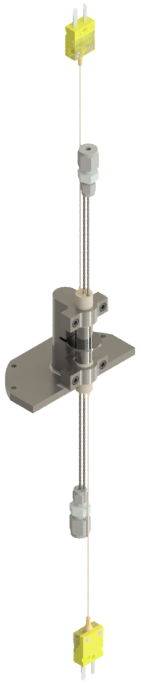}
         \put(17,58){\textcolor{black}{Electric}}
         \put(17,53){\textcolor{black}{Connections}}
         \end{overpic}
        
        \caption{Furnace views featuring four gas tubes (one inlet and one outlet in the top and bottom sides), two K-type thermocouples, and eight electric pin connections (four in the top and bottom sides). (a) Top view featuring the narrow X-ray entrance window, and the circular fluorescence window. (b) Internal view featuring the fuel cell holder assembly on the center.  }
        \label{designconcept}
\end{figure}

For in-situ X-ray absorption experiments, a key requirement is that the X-rays must reach the sample environment and both the fluorescence and the transmitted photons must be collected by X-ray detectors. This implies the inclusion of X-ray entrance and extraction windows in the sample furnace, as illustrated in figure \ref{designconcept}. In XAS it is also important to enable the collection of the X-ray fluorescence signal emitted by the sample. For this purpose, the furnace is equipped with a dedicated X-ray fluorescence window (see figure \ref{designconcept}).

The furnace will operate at the Quati beamline and will be placed on top of the beamline experimental table \cite{Santiago2023}. This setup will ensure mechanical support and allow the required movements in the X, Y, Z directions, as well as rotational movement ($\phi$ inclination).
As shown in figure \ref{designconcept} the furnace will feature a cell holder that is capable of securely hold the sample in the center position of the furnace. This will enable precise heating of the fuel cell pellet in a controlled environment and will help stabilize the temperature profile.

\begin{figure}
    \centering
    \includegraphics[width=.35\linewidth,trim=10.cm 0.cm 7.cm 0.cm, clip]{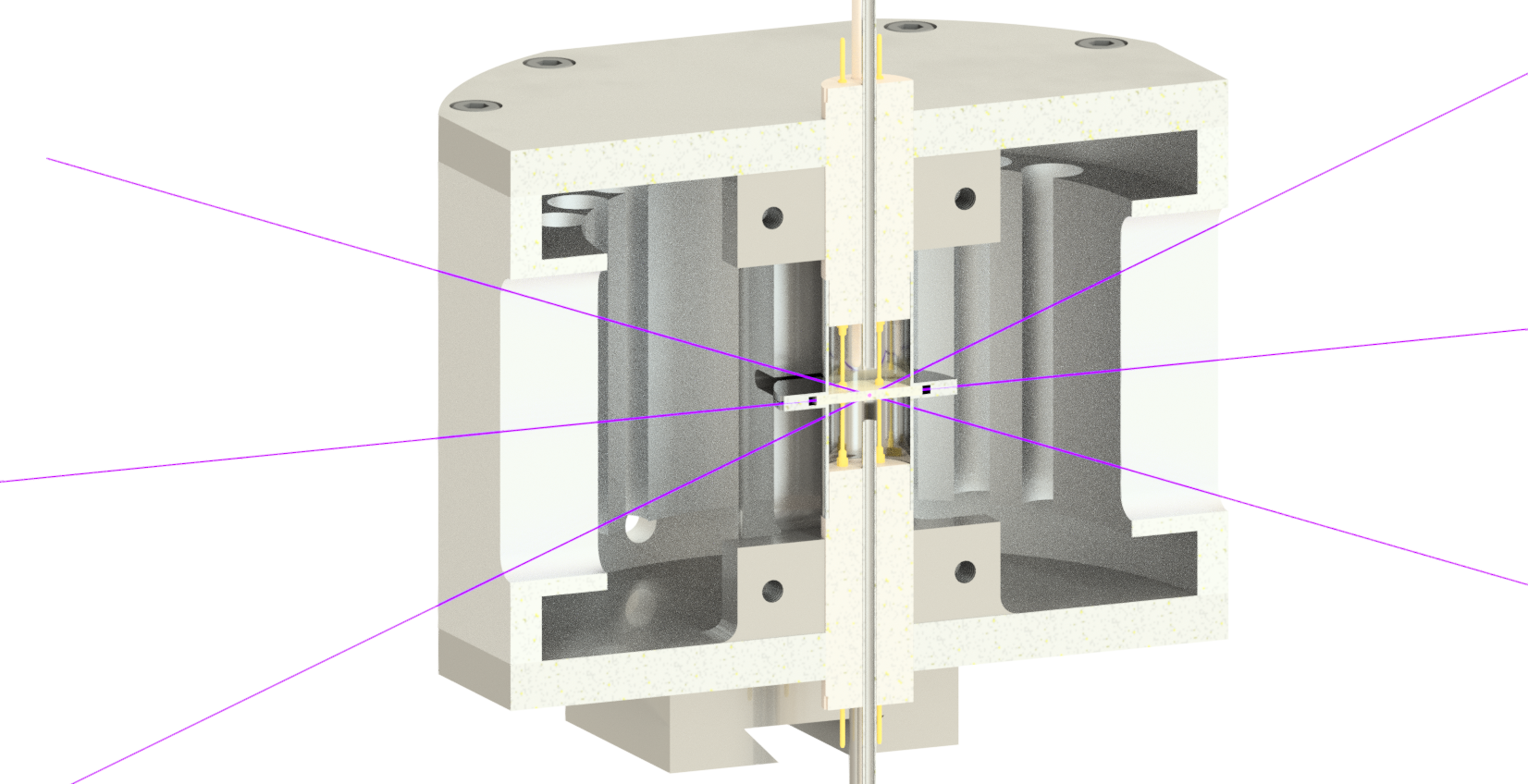}
    \caption{SOFC projection view showing the its internal parts and the X-ray beam path throughout the SOFC. In the figure the $\pm$20\textsuperscript{$\circ$}, and 0\textsuperscript{$\circ$} incident X-ray beams are represented by the purple lines. In the center, the gold plated contact pins (in yellow) are visible, and touching the SOFC pellet surface on both sides.}  
    \label{beampath}
\end{figure}

Heat will be provided by a 750W/220V heating resistor. This resistor will be routed around the fuel cell (see details on figure \ref{beampath}). The furnace will operate in air, and the windows will be covered with mica to minimize heat losses due to convection. In addition, cover plates will be placed on the top and bottom parts of the furnace to minimize heat losses by convection and improve temperature homogeneity and stability at 800\textsuperscript{$\circ$}C.

Inside the furnace, a fuel cell setup is placed. This device includes independent, interchangeable, and symmetric gas chambers positioned on opposite faces of the fuel cell pellet (see details on figure \ref{beampath}). This configuration allows one side of the fuel cell to be exposed to a reducing atmosphere, while the other side is exposed to an oxidizing gas. Differently from previous cell designs reported on the literature \cite{JieXiao2016,Skinner2013,Hanasaki2014,Francesco2019,Bryan2015,Seval2021}, in our design the symmetry of the gas chambers relative to the cell pellet enables switching gas types between the two chambers.

The gases used in the reactions will be conducted to the chambers using stainless steel gas tubes that are routed through an alumina cylindrical feedthrough, which provides mechanical stability. For each feedthrough, there is a gas inlet or injector positioned in the center, and a gas outlet on the side. The gas inlets will be positioned in a way that the gas is injected close to the fuel cell pellet surface, ensuring proper gas flow close to the surface.

Two K-type thermocouples are incorporated into the device to measure the temperature on both sides of the fuel cell pellet. In this design, the thermocouples are positioned close to the surface of the pellet to provide accurate temperature readings during in-situ studies. This configuration ensures that any temperature variations are closely monitored and taken into account in the analysis.

In order to close the electrical circuit and monitor the electrochemical reactions, the fuel cell is equipped with gold-plated electrical contact pins (see details on figure \ref{beampath}). These pins are spring-loaded to apply consistent pressure on the pellet surface, ensuring good electrical connectivity. Gold contacts are printed onto the cathode and anode surfaces to allow for efficient current flow. This configuration enables the transfer of electric charges from the anode - where the hydrogen oxidation reaction occurs - to the cathode, where electrons are consumed in the oxygen reduction reaction, sustaining the fuel cell operation.

A low impedance potentiostat is connected through a banana plug to alligator/crocodile clip cable to the fuel cell. It is used to apply voltage or current signals and monitor the system electrical response. By applying oscillating sinusoidal signals at different frequencies and amplitudes, it is possible to investigate the microscopic charge transport properties of the materials and measure the kinetics of fast electron transfer reactions. Our fuel cell configuration enables electrochemical analysis, as well as electrochemical impedance spectroscopy (EIS) measurements, a key technique for studying the dynamic behavior of the system.

In addition, due to the use of an X-ray beam to probe the fuel cell pellet surface, the fuel cell walls are made of boron nitride (BN) ceramic. The boron nitride structure is designed to offer both mechanical support and gas tightness, and at the same time allow the X-ray photons to reach the pellet during in-situ measurements.

The boron nitride material was chosen for the fuel cell wall due to its adequate mechanical properties and low X-ray absorption characteristics. This allows X-rays to reach the fuel cell pellet while maintaining the structural integrity of the device. The BN component is designed in a cylindrical shape with 300$\mu m$ thick wall. In this configuration, the beam path is such that more than 50$\%$ of the X-ray beam intensity – at beam energies above 7.1keV (Fe K edge) - reaches the surface of the pellet and penetrates the material (see figure \ref{beampath}). This ensures that sufficient X-ray photons reach the fuel cell pellet and probe the material.

To probe the fuel cell surface, the furnace features an X-ray entrance and extraction window that in combination with the fuel cell device geometry, allows the beam to access and collect information of surface dynamics by varying the incident X-ray beam angle. For the present design, the range of angular variation - from -20\textsuperscript{$\circ$} to 20\textsuperscript{$\circ$} (see figure \ref{beampath}) - makes it compatible with the grazing incident X-ray absorption spectroscopy (XAS) technique \cite{Seval2021,Yoichiro2020}.

Furthermore, to investigate the different interfaces and the bulk structure of the electrodes, control over the penetration depth of the X-rays is required. In XAS, this depth depends on both the X-ray energy and the angle of incidence between the incoming X-rays and the sample. Adjusting these parameters allows for depth-resolved studies of the fuel cell components \cite{Seval2021}. The combination of angle-dependent X-ray absorption techniques enables the acquisition of detailed structural and oxidation state information of the materials as a function of depth.

The furnace and fuel cell setup are designed to be fully integrated into the Quati beamline services \cite{Santiago2023}. This integration includes not only mechanical and optical alignment, but also compatibility with the gas injection system, electrical feedthroughs and connections, and X-ray beam path. 

Key requirements were considered in the design of the external hardware connections for the potentiostat. As mention previously, the electrical connection will be provided via eight banana plugs to alligator/crocodile clips. The cable routing follows the beamline layout while respecting current constraints related to positioning and safety. The cables connect the fuel cell device to a potentiostat (model BioLogic SP-200), which performs continuous measurements of impedance values during in-situ experiments. This setup allows precise monitoring of the cell electrochemical behavior in response to thermal and electrochemical changes across the sample.

To ensure stable operation under temperature, the system is designed to operate at a maximum temperature of 800\textsuperscript{$\circ$}C 
across the pellet surface. To reach and maintain this temperature, the setup integrates a 750W/220V heating resistor located around the fuel cell, along with two Omega K-type thermocouples. Both elements interface with an Eurotherm 3508 PID temperature controller, which receives feedback from the thermocouples. This configuration guarantees thermal stability during long-duration experimental runs.

To prevent short circuit, avoid gas leaks around the electrodes, and ensure gas tightness, the fuel pellet is attached to the boron nitride gas chamber using insulating ceramic sealant agent (PERMATEX 80335 Sealer) as both a sealing and jointing material. The same ceramic adhesive is used to secure the electrical contacts, thermocouples, and gas tubes in their positions.

\begin{figure}
\centering
    \includegraphics[width=0.35\linewidth,trim=1.cm 5.cm 2.cm 8.5cm, clip]{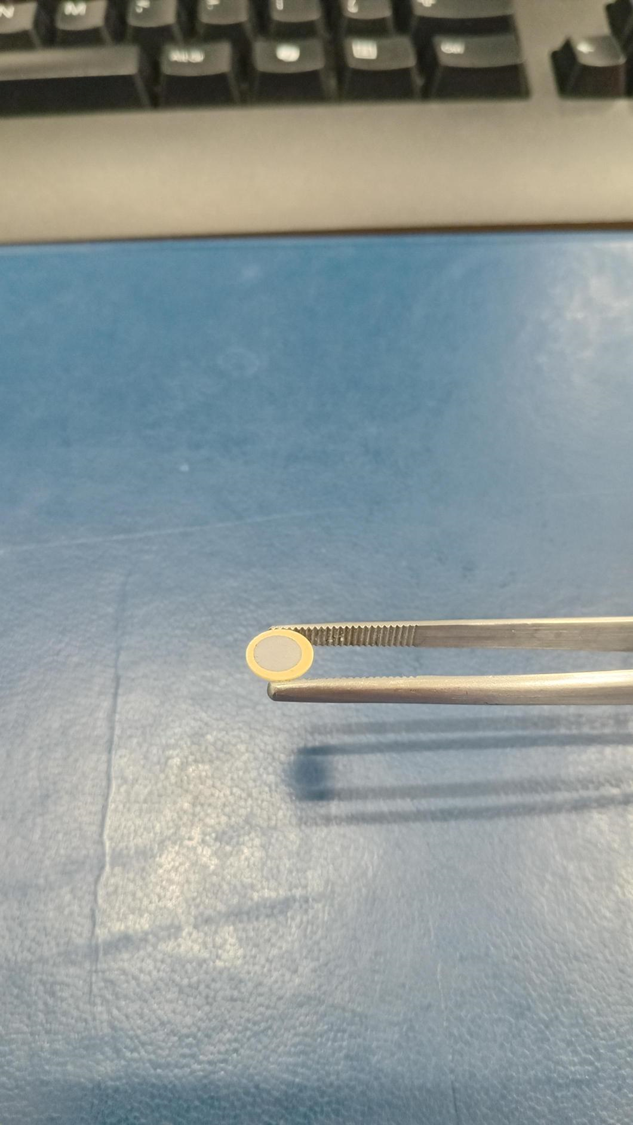}
    \caption{9$mm$ diameter and 600$\mu m$ thick GDC ceramic pellet with circular platinum (Pt) electrodes. The pellet were sintered in air at 1300 \textsuperscript{$\circ$}C for four hours. Circular platinum electrodes with a diameter of 6 $mm$ and approximately 0.2 $\mu m$ thick were deposited in the center on both faces of the pellet by electron‑beam deposition.}
    \label{pellet}
\end{figure}

\section{Results}

Proof of concept electrochemical impedance spectroscopy measurements were carried out in air over a temperature range from 200\textsuperscript{$\circ$}C to 350\textsuperscript{$\circ$}C, for gadolinium doped ceria (GDC) ceramic pellets produced in-house (see figure \ref{pellet}). The pellet was accommodated accordingly inside a tubular furnace, and the contacts established with a nickel foam collectors. The electrochemical impedance spectra were obtained using the potentiostatic electrochemical impedance spectroscopy \cite{BioLogic} technique, applying a sine wave as the excitation signal with frequencies ranging from 1MHz to 1Hz, with an amplitude of 100mV and under open circuit voltage conditions.

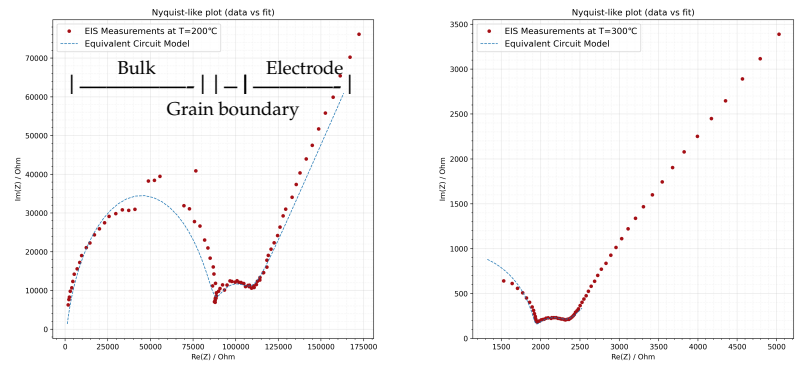
\begin{figure}
    \centering
    \begin{overpic}[width=.35\linewidth]{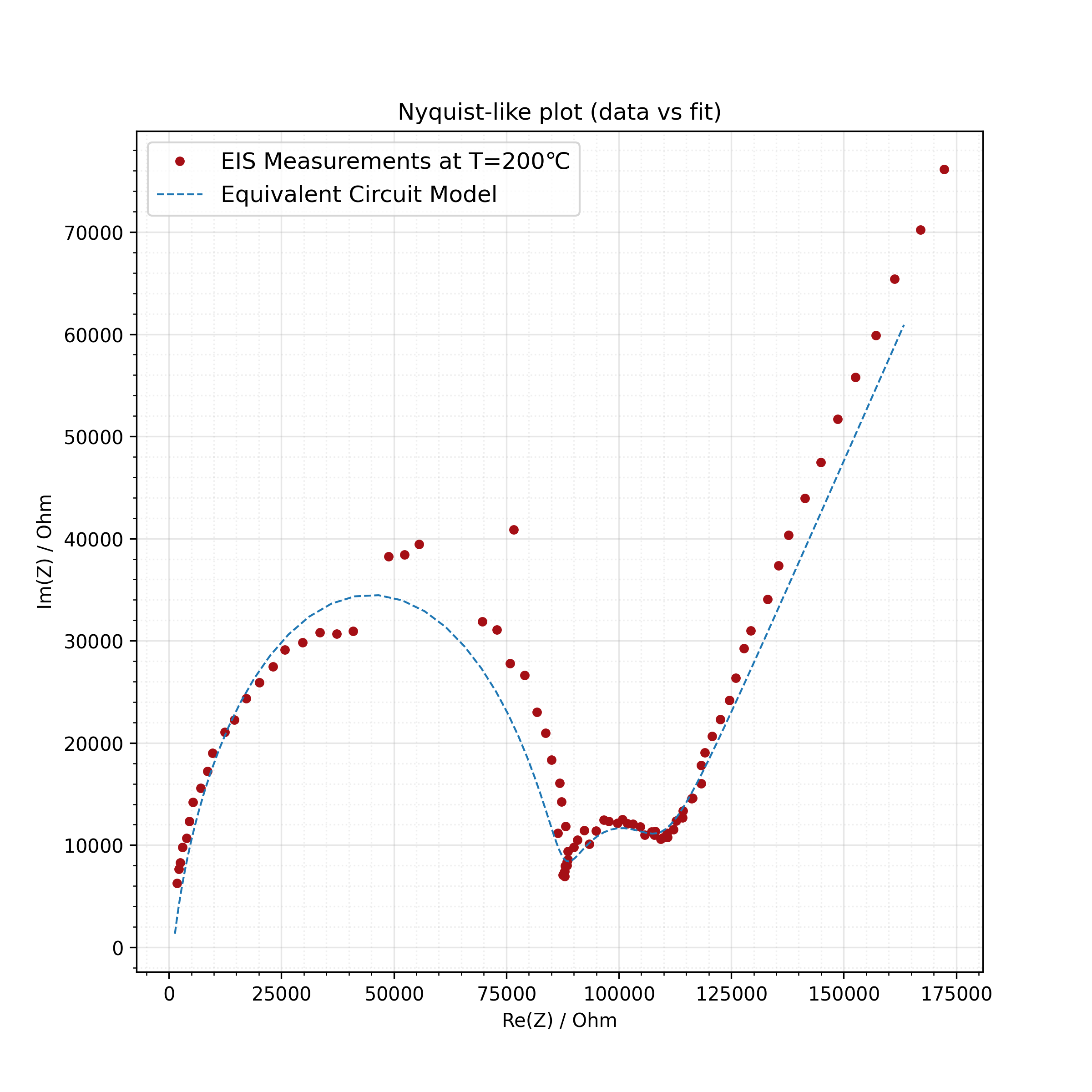}
    
    \put(28,75){\textcolor{black}{\scriptsize Bulk}}
    \put(40,65){\textcolor{black}{\scriptsize Grain boundary}}
    \put(64,75){\textcolor{black}{\scriptsize Electrode}}

    \put(15,70){\textcolor{black}{|-------------|}}
    \put(50,70){\textcolor{black}{|--|}}
    \put(57,70){\textcolor{black}{|----------|}}
    \end{overpic}
    \includegraphics[width=.35\linewidth]{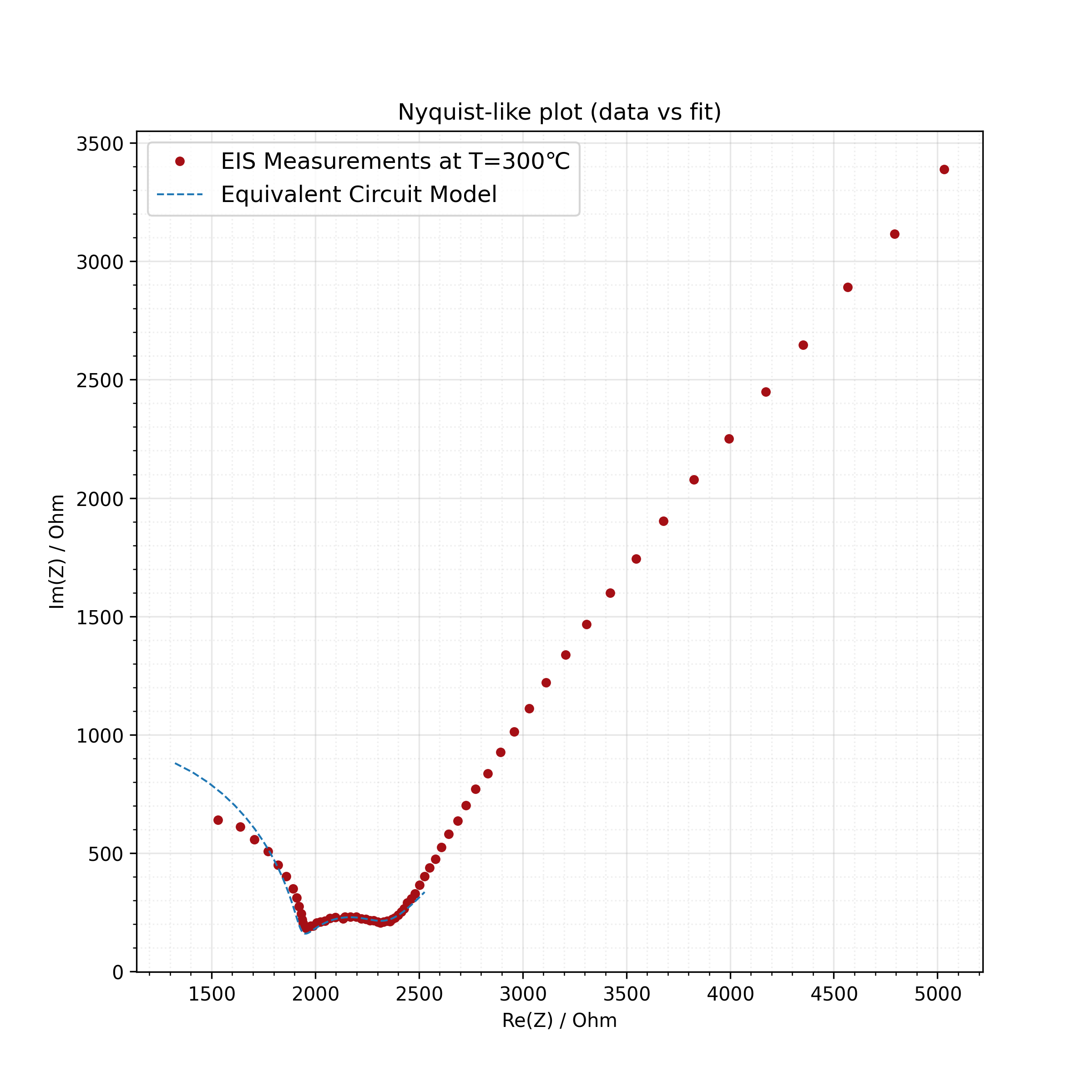}
    \caption{Nyquist plots for T=200\textsuperscript{$\circ$}C and 300\textsuperscript{$\circ$}C. Both plots shown two partially overlapping semicircular arcs followed by a low-frequency tail related to electrode contribution.}
    \label{impedance_results}
\end{figure}

Nyquist plots exhibit the characteristic response of dense oxygen-ion conducting ceramics, consisting of two partially overlapping semicircular arcs followed by a low-frequency electrode-related contribution (see figure \ref{impedance_results}). The high-frequency semi-circle corresponds to the bulk resistivity, whereas the intermediate-frequency arc is attributed to grain-boundary contributions. The low frequency tail, is associated with electrode polarization.

The EIS data were acquired and directly stored in the Sirius data center for subsequent analysis, demonstrating the beamline capability to perform real-time operando electrochemical characterization under controlled temperature conditions. In combination with X-ray absorption spectroscopy (XAS) measurements, the datasets are synchronized via the beamline trigger unit, enabling direct correlation between the EIS response and the XAS spectral information.

\begin{figure}[h]
\centering
    \includegraphics[width=0.35\linewidth,trim=0.cm 0.cm 0.cm 0.cm, clip]{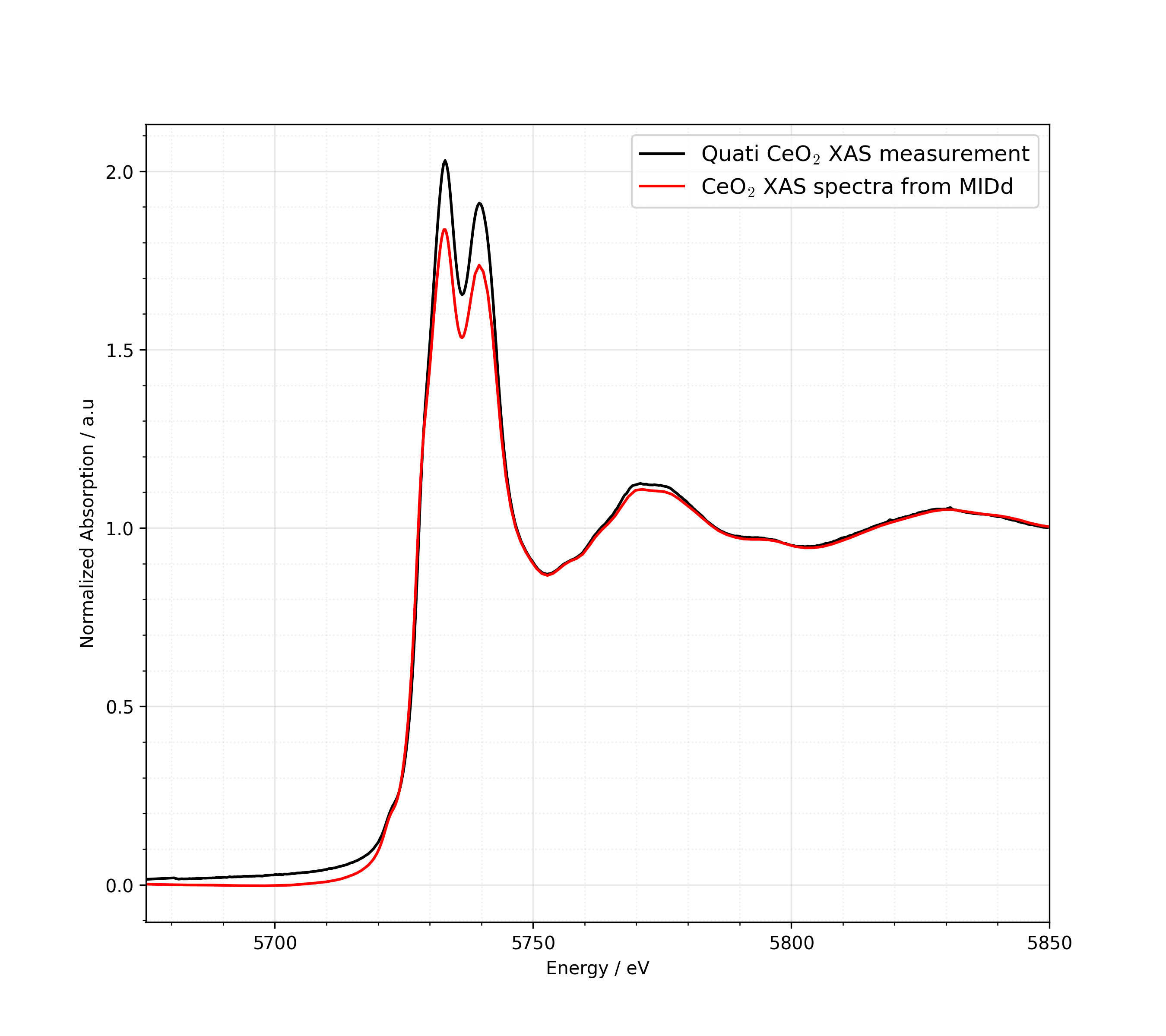}
    \caption{Room temperature Ce L3 edge XANES spectra. The XAS spectra in red was reproduced with permission from \cite{MIDb}.}
    \label{xas}
\end{figure}

In addition to electrochemical characterization, separated standalone L3 edge XANES  measurements for ceria oxide sample (CeO$_2$) were performed in transmission mode and room temperature conditions. Figure \ref{xas} shows the normalized XANES spectra for CeO$_2$ which is characterized by the presence of a double-peak shape L3 edge \cite{Fonda1999EXAFSAO}. In the plot, the black line represents the measurement performed at Quati beamline, and in red the XAS spectra available in the literature \cite{MIDb} highlighting the agreement between both measurements.

Within this frame work the advantages of EIS and XAS combined are leveraged to access information regarding the local atomic structure and electronic configuration under realistic electrochemical conditions \cite{CABRERAPASCA2026}, with high temporal and spatial resolution. Throughout the integration of EIS and X-ray absorption spectroscopy analysis, we expect have detailed information in order to tackle the issues regarding the long-term durability of SOFC devices.


\section{Conclusions}

In this work the details of a new sample cell for in-situ/operando studies of symmetric IT-SOFC are presented and discussed. The proposed electrochemical sample cell architecture is expected to consolidate EIS and synchrotron based XAS as primary techniques for providing information regarding  alterations in the local atomic structure and oxidation state under realistic electrochemical operation conditions.   

Electrical impedance spectroscopy of gadolinia-doped ceria electrolytes were measured in air using in-house developed instrumentation and software tools to control the experiment, handle the data acquisition, and the data analysis. The impedance spectrum analysis reveal the presence of two distinct contributions, one from the bulk material resistivity, and another from the grain boundary resistivity.

Standalone L3 edge XANES measurement for CeO$_2$ were performed in transmission mode showing the agreement with data available in the literature \cite{MIDb}. This measurement will allow further in-situ/operando studies of catalytic compounds containing tetravalent cerium such as gadolinia-doped ceria, as well as other materials.

The preliminary results suggest a potential use of this type of
cell, and both techniques, for the investigation of working electrodes for IT-SOFC applications.

\vspace{60pt}

\authorcontributions{.}

\funding{FAPESP process 24/01511-3 and 24/20599-9.}

\institutionalreview{.}

\informedconsent{.}

\dataavailability{.} 

\acknowledgments{The authors would like to thank the São Paulo Research Foundation (FAPESP) for its financial support (process 24/01511-3 and 24/20599-9).}

\conflictsofinterest{.}

\isPreprints{}{
} 

\reftitle{References}


\bibliography{mybib}

%


\isPreprints{}{
} 
\end{document}